\documentclass[reqno]{amsart}
\usepackage{amscd}
\usepackage[dvips]{graphics}

\def\beq{\begin{equation}}
\def\eeq{\end{equation}}
\def\ba{\begin{array}}
\def\ea{\end{array}}

\numberwithin{equation}{section}
\newenvironment{abs}{\textbf{Abstract}\mbox{  }}{ }
\newenvironment{key words}{\textbf{Keywords}\mbox{  }}{ }
\newtheorem{theorem}{Theorem}[section]
\newtheorem{definition}[theorem]{\textbf{Definition}}

\renewenvironment{proof}{\noindent{\textbf{Proof.}}}{\hfill$\Box$}
\theoremstyle{remark}
\newtheorem{remark}[theorem]{\textbf{Remark}}
\theoremstyle{plain}

\begin{document}
\title{\textbf{On the sparsity of binary numbers}}
\author  {Meijun Zhu}
\address{Department of Mathematics, The University of Oklahoma, Norman, OK 73019, USA; Everyone Academy, 27 Hancock Ave. Ext. \#B Medford, MA 02155, https://everyone-math.com/}

\email{mzhu@math.ou.edu, mthcnt@gmail.com}

%\date{10-08-2024}
%\date{10-14-2024 to Arxiv}
\maketitle

% ----------------------------------------------------------------
\noindent
\begin{abs}
 We introduce the concept of negative coefficients in various number-based systems, with a focus on decimal and binary systems. We demonstrate that every binary number can be transformed into a sparse form, significantly enhancing computational speed by converting binary numbers into this form.
\end{abs}\\

\begin{key words} Decimal number, Binary number, Negative coefficient, Sparsity, Computation speed
\end{key words}\\
\textbf{Mathematics Subject Classification(2020).}
90C09, 90C10, 97H20 \indent
%---------------------------------------------------------------------------------
\section{\textbf{Introduction and motivation}\label{Section 1}}

 One challenge in teaching college math courses is that many students struggle with polynomial long division, even though they are proficient with long division for decimal numbers. To bridge the gap between these two seemingly ``different'' types of division, we introduced negative coefficient notations for decimal numbers in  \cite{ZW}. First, we define the opposite of a number $A$ as $\hat{A}$, allowing all negative numbers to be expressed, for example, as:
$$
\hat 1=-1, \  \hat 8=-8, \  \hat{10}=-10, \ \ \hat{18}=\hat 1  \hat 8=-18.
$$

This allows us to represent decimal numbers in an alternative form. For example, we can express
$$
4\times 10^2 - 2 \time 10+1 \times 1= 4\hat 2 1
$$
which in fact, is $381$.

The advantage of this notation is that it reveals cancellations during calculations. For instance:
$$4 \hat 2 1 \times 21=8001.$$
The vertical form provides a clearer view of the calculation, as shown here:

 \[ \begin{array}{*5r}
      &     & 4   &\hat 2   &1\\
\times&     &     &2    &1\\ \hline
     &      & 4     &\hat 2    &1\\
    +&8     &\hat 4      &2    &\\ \hline
     &8      &0      &0     &\ 1.\\
   \end{array} \]

By using negative coefficients, the connection between the product of decimal numbers and polynomial multiplication becomes evident. For instance, from $2 \hat 3 \times 31=6\hat 7 \hat 3$, one can easily see the polynomial equivalent: $(2x-3)(3x+1)=6x^2-7x-3,$ and vice verse.

\medskip

This approach also simplifies division. For instance, when calculating $27001\div 31$, the traditional long division can be performed as the following:
\[
\renewcommand\arraystretch{1.2}
\begin{array}{*1r @{\hskip\arraycolsep}c@{\hskip\arraycolsep} *6r}
 &&  &   &   8 &	7&\bf{1} &\\
\cline{2-7}
31&\big) &2&7 & 0 & 0 & 1 \\
 &-& 2 & 4  & 8	  &	& &\\
\cline{3-7}
 &&  & 2 & 2	  &	0& &\\
 && - & 2 & 1	  &	7& &\\
\cline{3-7}
  &&  &  & 	  &	3&1 &\\
  &&  &  & 	  -&	\bf{3}&\bf{1} &\\
\cline{3-7}
&&  &  & 	  &	&\bf{0}. &\\
\end{array}
\]

However, using negative coefficients, the operation becomes more straightforward, as illustrated here:

\[
\renewcommand\arraystretch{1.2}
\begin{array}{*1r @{\hskip\arraycolsep}c@{\hskip\arraycolsep} *6r}
 &&  &   &   9 &	\hat 3&\bf{1} &\\
\cline{2-7}
31&\big) &2&7 & 0 & 0 & 1 \\
 &-& 2 & 7  & 9	  &	& &\\
\cline{3-7}
 &&  &  & \hat 9	  &	 0& &\\
 && - & & \hat 9	  &	 \hat 3 & &\\
\cline{3-7}
  &&  &  & 	  &	3&1 &\\
  &&  &  & 	  -&	\bf{3}&\bf{1} &\\
\cline{3-7}
&&  &  & 	  &	&\bf{0}. &\\
\end{array}
\]

This is almost identical to the polynomial division $(27x^3+1)\div (3x+1)$, shown below:

\[
\renewcommand\arraystretch{1.2}
\begin{array}{*2r @{\hskip\arraycolsep}c@{\hskip\arraycolsep} *4r}
  &    &&    &  9x^2 &    -3x &   +1 \\
\cline{3-7}
3x  +&1 &\big)& 27x^3 &+0x^2 &+0x&+ 1 \\
  &    && -( 27x^3 & + 9x^2)& \\
\cline{4-7}
  &    &&    &-9x^2&+0x \\
  &    &&   &-(-9x^2&-3x)& \\
\cline{5-7}
  &    &&    & &3x & +1\\
  &    &&    &    &   -(3x &+ 1) \\
\cline{6-7}
 &    &&    &    &    & 0. \\
\end{array}
\]

While the new notation for decimal numbers (using a hat to indicate negative coefficients) aims to enhance the efficiency of math education, we discover that applying negative coefficients to binary numbers has the potential to significantly improve computation speed and may have broader applications.

\section{Main result: Sparsity on binary numbers}

Recall a binary number is give in the form
$$
c_n\times 2^n+ c_{n-1} \times 2^{n-1}+ \cdots... +c_1 \times 2+c_0,
$$
where $c_n=1$, $c_i=0 $ or $1$ for $i=0, 1, \cdots, n-1.$	We typically use subscript $2$ to indicate that a number is in binary form. For example:
$$
1011_2=1\times 2^3+0 \times 2^2+1 \times 2 +1.
$$
With the introduction of negative coefficient, we can alternatively express this as
$$
1011_2=110\hat 1_2.
$$

We now introduce the concept of the sparse form for binary numbers.

\begin{definition}A binary number is in sparse form if there are no consecutive positions where the coefficients are non-zero.
\end{definition}

For example,
$
10010101_2
$
is in sparse form, whereas
$
100110_2
$ is not.

\medskip

We first observe that a non-sparse binary number can be converted into sparse form.  For example:
$$
100110_2=1010\hat 10_2.
$$

The main result we aim to establish in this note is the following discovery:

\begin{theorem}
Every binary number has a unique sparse form.
\end{theorem}

\begin{proof}
To convert a non-sparse binary number to its sparse form, we begin with the ones and tens places.
Without loss of generality, assume the number in the ones place is not zero (if the ones place is zero, we proceed by considering the tens and hundreds places, and apply induction to obtain the result).
If the tens place is zero, the binary number is already in sparse form for the ones and tens places, and we can move on to the hundreds and thousands places. By continuing this process and using induction, we achieve the desired sparse form.

\smallskip

Otherwise,

\noindent If it is in the form of ``$\cdots 111_2$", then it can be converted to ``$\cdots 00\hat 1$".

\smallskip

\noindent If it is in the form of ``$\cdots 011_2$", then it can be converted to ``$\cdots 10\hat 1$".

\smallskip

\noindent If it is in the form of ``$\cdots \hat 1 11_2$", then it can be converted to ``$\cdots 00\hat 1$".

\smallskip

\noindent If it is in the form of ``$\cdots 1\hat 1_2$'', then it can be converted to ``$\cdots 0 1$".

\smallskip

\noindent If it is in the form of ``$\cdots \hat 11_2 $'', then it can be converted to ``$\cdots 0 \hat 1$".

\smallskip

And also easy to see (using symmetric property):

\noindent If it is in the form of ``$\cdots 1\hat 1 \hat 1_2$", then it can be converted to ``$\cdots 00 1$".

\smallskip

\noindent If it is in the form of ``$\cdots 0 \hat 1 \hat 1_2$", then it can be converted to ``$\cdots \hat 10 1$".

\smallskip

\noindent If it is in the form of ``$\cdots \hat 1 \hat 1 \hat 1_2$", then it can be converted to ``$\cdots 00 1$".

\smallskip

In all cases, we can first convert the number into its sparse form for the ones and tens places, and then proceed to the hundreds and thousands places. By continuing this process and applying induction, we eventually obtain the complete sparse form of the binary number.

\medskip

To demonstrate that the sparse form is unique, we can use the proof by contradiction.

\medskip

Suppose a binary number has two different sparse forms, $A$ and $B$, which differ at the ten-thousands place. If they differ at a place higher than the ten-thousands, the argument remains the same. If they differ at a place lower than the ten-thousands, we can artificially multiply both numbers by $2^n$  to shift them to the ten-thousands place.

 Without loss of generality, assume the digit in the ten-thousands place for $A$ is $1$, while the digit in the ten-thousands place for $B$ is different. In this case, it must be $0$; otherwise, we would have $A-B>0$. Therefore, we can express them as follows (recall that they are in sparse form):

\begin{align*}
A&=......10xyz_2\\
B&=......010uv_2.
\end{align*}

Now, the smallest value that $x$ could be is $\hat 1$ (which implies that $y$ must be $0$ since $A$ is in sparse form). The largest value that $u$ could take is $1$ (which means that $v$ must be $0$), and the smallest value that $z$ could be is $\hat 1$. Therefore
$$
A-B \ge 1\hat 1 \hat 1 \hat 1=1,
$$
which contradicts the fact that $A=B$. Thus, the proof is complete.

\end{proof}

\begin{remark}
The proof of the theorem also outlines the algorithm for converting a binary number into its sparse form.
\end{remark}

\section{Conclusion and outlook}

The computation speed for addition and multiplication of binary numbers in sparse form is clearly faster than that for non-sparse form. Here is one typical example for the addition:

\[ \begin{array}{*9r}
    & &  1 & 0     &\hat 1  & 0  &0 & 1 & 0 \\
+  & &  1  & 0     &\hat 1 & 0 & \hat 1 & 0 & 1 \\ \hline
   &1 & 0 & \hat 1 & 0  &0      &\hat 1     &1     &\ 1\\
   \end{array} \]

In this representation, the processes of borrowing and carrying numbers are essentially eliminated, simplifying the computation.

Further experimental research will be conducted in the future. The sparsity of other number-based systems and their applications, among other topics, continues to present theoretical challenges.

 \vskip 1cm
 \noindent {\bf Acknowledgements}\\
 \noindent  We would like to express our gratitude to Helen Wu for being our first audience and for her encouragement to share this note with a broader audience.
 %This work is trademarked and patented by the author and Helen Wu.
 %% bibliography--------------------------------------------------------------------
%\begin{center}

\end{document}